\documentclass[aps,nofootinbib,superscriptaddress, showpacs,preprintnumbers,  nofootinbibt,twocolumn]{revtex4-2}

\usepackage[utf8]{inputenc}
\usepackage[T1]{fontenc}
\usepackage[english]{babel}
\usepackage{soul}
\usepackage{amsmath, amssymb, amsfonts, bm}
\usepackage{mathtools} 
\usepackage{physics} 
\usepackage{siunitx} 

\usepackage{graphicx}
\usepackage[justification=justified]{caption}
\usepackage{ragged2e}
\usepackage{lipsum}

\usepackage{float} 
\usepackage{subcaption} 

\usepackage{tikz}
\usetikzlibrary{arrows.meta, calc, decorations.markings}
\tikzset{short/.style={thick}}
\usetikzlibrary{quotes,angles}

\usepackage{booktabs}
\usepackage{multirow}
\usepackage{siunitx}

\usepackage{xcolor}
\usepackage{hyperref}
\hypersetup{
 colorlinks=true,
 linkcolor=blue,
 citecolor=blue,
 urlcolor=magenta,
}

\newcommand{\beqn}{\begin{eqnarray}}
\newcommand{\eeqn}{\end{eqnarray}}

\usepackage{xcolor}

\begin{document}

\title{Probing Dark Matter and Phantom Field Effects on Neutrino Oscillations around Black Holes}

\author{Ikrom Ergashov}
\email{i.ergashov@newuu.uz}
 \affiliation{Institute for Advanced Studies, New Uzbekistan University, Movarounnahr str. 1, Tashkent 100000, Uzbekistan}

\author{Bakhtiyor Narzilloev}
	\email{b.narzilloev@newuu.uz}
 \affiliation{Institute for Advanced Studies, New Uzbekistan University, Movarounnahr str. 1, Tashkent 100000, Uzbekistan}

\author{Ibrar Hussain}
	\email{ibrar.hussain@seecs.nust.edu.pk}	
	\affiliation{School of Electrical Engineering and Computer Science, National University of Sciences and Technology, H-12, Islamabad, Pakistan}
    \affiliation{Research Center of Astrophysics and Cosmology, Khazar University, 41 Mehseti Street, AZ1096 Baku, Azerbaijan}

\author{Bobomurat Ahmedov}
	\email{ahmedov@astrin.uz}
 \affiliation{Institute for Advanced Studies, New Uzbekistan University, Movarounnahr str. 1, Tashkent 100000, Uzbekistan}
 \affiliation{School of Physics, Harbin Institute of Technology, Harbin 150001, People’s Republic of China}
 \affiliation{Institute of Theoretical Physics, National University of Uzbekistan, Tashkent 100174, Uzbekistan}

\begin{abstract}
This study investigates how dark matter in the background of the phantom field (DMPF) near a black hole influences neutrino flavor oscillations using a two-flavor model. It is found that gravitational effects are cancelled for radially moving neutrinos, causing the oscillation phase to increase with distance and mass-splitting, similar to flat spacetime. However, deflected neutrinos experience additional phase shifts due to dark matter, which slightly alters the mass-difference term and the total phase. Numerical analysis supports these findings and reveals a degeneracy between lens mass and dark matter parameters, affecting the transition probability curves based on source angle changes. Neutrinos modeled as Gaussian wave packets show that decoherence in a gravitational field is minimally influenced by dark matter, with the absolute neutrino mass determining the oscillation duration. Our findings imply that curved spacetime and dark matter significantly impact neutrino oscillations, offering potential insights into dark matter through neutrino astronomy in extreme environments. 
\end{abstract}

\maketitle

\section{Introduction}

Black hole solutions represent some of the most profound and intriguing predictions of Einstein’s theory of General Relativity (GR), and they have played a central role in testing gravity in the strong-field regime. In particular, the landmark detection of gravitational waves from black hole mergers \cite{1a} and the direct imaging of supermassive black holes at the centers of galaxies, such as M87 and the Milky Way \cite{3a,4a}, have provided compelling observational confirmation of GR under extreme conditions.

Modern cosmological observations reveal that the universe is undergoing accelerated expansion \cite{4ab}. Within the framework of the Standard Model of Cosmology (SMC), this acceleration is commonly attributed to dark energy, which constitutes about 68\% of the total cosmic energy budget, while approximately 27\% is dark matter and only about 4\% is ordinary baryonic matter \cite{5a,6a}. These findings naturally motivate the investigation of black hole solutions in environments where dark matter or dark energy is present. In this regard, black holes surrounded by various forms of dark energy have received significant attention, and both static and rotating configurations have been explored in the contexts of particle dynamics \cite{7a,8a,9a,10a,11a,12a} and thermodynamics \cite{13a,14a,15a}. Among the proposed dark matter candidates, perfect fluid dark matter, introduced by Kiselev and later developed by Li and Yang \cite{16a}, provides a reasonable description of the nearly flat rotation curves observed in spiral galaxies \cite{17a,18a,19a}. Recognizing the importance of DMPF, the Schwarzschild black hole solution has been extended to include DMPF within the Einstein field equations \cite{20a}, and this modified geometry has since been studied in several physical contexts (see, for example, \cite{21a,22a,23a,24a,25a}).

The study of neutrino oscillations on curved background is of significant importance in astrophysics and astroparticle physics \cite{5b,5c,5d}. Particularly in extreme environments such as core-collapse supernovae or regions surrounding compact massive objects like black holes, spacetime curvature can markedly influence neutrino propagation, and analyzing these effects provides valuable insights into the fundamental properties of neutrinos as well as their interaction with gravity \cite{5b,5e,6b,7b,8b}. These investigations can deepen our understanding of both particle physics and GR. Any departure from the predictions of GR may provide compelling evidence in favor of modified gravity, highlighting the importance of a comprehensive, multi-pronged strategy for testing gravitational theories across diverse physical systems and length scales, as demonstrated in numerous theoretical investigations~\cite{C3,C4,C5,Narzilloev20b, Narzilloev21b, Narzilloev22, Narzilloev22b, Narzilloev2023b, Narzilloev2023d, Narzilloev2023e, Narzilloev2024b}. In this work, we focus on examining how the presence of DMPF near the Schwarzschild black hole affects neutrino flavor oscillations.

In this study, in the presence of DMPF  near a neutral static black hole, we consider the scenario of the neutrino flavor oscillations within a two-flavor oscillation model. Our findings show that neutrinos moving directly outward from the black hole do not experience net gravitational phase contributions because the relevant gravitational terms cancel. As a result, the oscillation phase grows with the propagation distance and the mass-squared difference exactly as it does in flat spacetime. The situation changes for neutrinos whose paths are bent by the gravitational field. These deflected neutrinos accumulate extra phase contributions that arise from the interaction between spacetime curvature and DMPF. This subtly modifies the effective mass-splitting parameter and shifts the overall oscillation phase. Numerical calculations reinforce these results and reveal a degeneracy between the mass of the lensing object and the dark sector parameter, which manifests itself as noticeable changes in the transition probability curves when the emission angle of the neutrino source is varied. Treating neutrinos as Gaussian wave packets, we further show that the decoherence effects in curved spacetime are only minimally influenced by the surrounding DMPF . Instead, the absolute neutrino mass plays the dominant role in determining the coherence length and the time over which oscillations persist. These outcomes indicate that both the geometry of curved spacetime and the presence of dark matter can meaningfully modify neutrino oscillation behavior, suggesting a promising way to probe dark matter distributions using neutrino signals from extreme astrophysical environments.

The paper is organized as follows: In the next Section we discuss the metric of the Schwarzschild black hole in the presence of the DMPF. Section III is devoted to neutrino oscillations in Minkowski spacetime. In Section IV, we analyze the phase difference on the curved background. In Section V, the radial and non-radial propagation of neutrinos in the equatorial plane is considered. Neutrino oscillation probabilities are discussed in Section VI. In Section VII the two-flavor approximation model for neutrino oscillation is studied. Neutrion decoharence is presented in Section VIII. Conclusion and discussion of the work done is given in the last Section.   

\section{Structure of the Spacetime Metric}
Before we begin with the main problem, let us briefly review the spacetime of the black hole surrounded by dark matter existing within a phantom field background. We shall begin with the Einstein-Hilbert action modified with a scalar field to model the dark-matter component. The action which incorporates both gravitational and scalar field contributions is given by~\cite{Kiselev2003, Li2012}:
\begin{eqnarray}
S &=& \int d^{4}x\, \sqrt{-g}\nonumber \\ &&\times \left[\frac{R}{2\kappa^{2}} + \frac{1}{2} g^{\mu\nu}\partial_{\mu}\Phi\, \partial_{\nu}\Phi - V(\Phi) + \mathcal{L}_{m} + \mathcal{L}_{I}\right],
\end{eqnarray}
where $R$ is the Ricci scalar representing the curvature of spacetime, $\kappa^{2} = 8\pi G$ (with $G$ being the gravitational constant), $\Phi$ is a scalar field associated with the dark matter and  phantom field, and $V(\Phi)$ is its potential, which rises due to the self-interaction of the scalar field. The term $\mathcal{L}_{m}$ denotes the Lagrangian of standard matter fields or baryonic matter, while $\mathcal{L}_{I}$ is for possible interactions between the dark matter scalar field and baryonic matter, allowing for a more general coupling that could influence the dynamics of the system.

Varying the action with respect to the metric $g_{\mu\nu}$ yields in the Einstein field equations:
\begin{equation}
G_{\mu\nu} = \kappa^{2} \left( T_{\mu\nu}^{\Phi} + T_{\mu\nu}^{m} + T_{\mu\nu}^{I} \right), 
\end{equation}
here, $G_{\mu\nu}$ is the Einstein tensor along the energy-momentum tensors $T_{\mu\nu}^{\Phi}$ for the scalar field along with $T_{\mu\nu}^{m}$ and $T_{\mu\nu}^{I}$ which stand for standard matter and interaction terms, respectively. The energy–momentum tensor corresponding to the scalar field is expressed as
\begin{equation}
T_{\mu\nu}^{\Phi} = \partial_{\mu}\Phi \partial_{\nu}\Phi - \frac{1}{2} g_{\mu\nu} \partial^{\lambda}\Phi \partial_{\lambda}\Phi + g_{\mu\nu} V(\Phi). 
\end{equation}

The spacetime geometry corresponding to a black hole immersed in DMPF  can be expressed through the following line element, as presented in Refs.~\cite{Li2012, Shaymatov2021}. It is easier and more convenient to use line element as $ ds^2 = -\mathcal{A} dt^2 + \mathcal{B} dr^2 + \mathcal{C} d\theta^2 + \mathcal{D} d\phi^2 $, where 
\begin{align}
\mathcal{A}(r) &= f(r) = 1 - \frac{2M}{r} + \frac{a}{r} \ln\left(\frac{r}{|a|}\right),\label{eq:7} \\ 
\mathcal{B}(r) &= \frac{1}{f(r)} = \frac{1}{1 - \frac{2M}{r} + \frac{a}{r} \ln\left(\frac{r}{|a|}\right)},\label{eq:8}\\
\mathcal{C}(r) &= r^{2}, \\
\mathcal{D}(r) &= r^{2} \sin^{2}\theta.
\end{align}

This metric represents a static and spherically symmetric black hole solution, where the function $f(r)$ encapsulates the influence of both the central mass and the surrounding dark matter distribution, and 
here, $M$ is the mass of the black hole, and the logarithmic term, proportional to the parameter $a$, arises from the contribution of the scalar field.  The parameter $a$, often termed the "dark matter density parameter", represents the entire dark sector (dark matter and dark energy), with components differing by a constant ratio, making it more accurately a dark sector density parameter. The total energy density and dark matter energy density are given by
\begin{align}
\rho_{\rm total} = \frac{a}{8\pi r^{3}},\quad \rho_{\rm DM} = \frac{3a}{16\pi r^{3}}.
\end{align}%

\section{Neutrino Oscillations in Flat Spacetime}
Neutrinos are fundamental particles in the Standard Model, and can be produced and detected in flavor states, which are indicated by $|\nu_{\alpha}\rangle$, where $\alpha=e, \mu, \tau$ corresponds to electron, muon, and tau neutrinos. These flavor states are not Hamiltonian eigenstates, but are superpositions of the mass eigenstates $|\nu_{i}\rangle$ (where $i=1, 2, 3$), related by the $3\times 3$ Pontecorvo-Maki-Nakagawa-Sakata (PMNS) \cite{Pontecorvo1958, Pontecorvo1968, Maki1962} leptonic mixing matrix. The relationship can be expressed as
\begin{equation}
|\nu_{\alpha}\rangle=\sum_{i=1}^{3}U^{*}_{\alpha i}\, |\nu_{i}\rangle, 
\end{equation}
where $U_{\alpha i}$ are the elements of the PMNS matrix, which encodes the mixing angles and responsible for the CP violation.

Let us consider a neutrino originating from a source $S$, located at spacetime coordinates $(t_{S}, x_{S})$, and leter observed at a detector $D$, positioned at $(t_{D}, x_{D})$. The progression of a mass eigenstate $|\nu_{i}\rangle$ as it passes through flat spacetime is characterized by a plane-wave formulation:
\begin{equation}
|\nu_{i}\left(t_{D}, x_{D}\right)\rangle=\exp\left(-i\Phi_{i}\right)|\nu_{i} \left(t_{S}, x_{S}\right)\rangle, 
\end{equation}
where $\Phi_{i}$ represents the phase accumulated by the $i$-th mass eigenstate during its propagation. The probability of a neutrino produced in flavor state $|\nu_{\alpha}\rangle$ being detected in flavor state $|\nu_{\beta}\rangle$ is determined by:
\begin{eqnarray}
P_{\alpha\beta}&=&|\langle\nu_{\beta}|\nu_{\alpha}\left(t_{D}, x_{D}\right)\rangle|^{2}\nonumber \\ &=&\sum_{i, j=1}^{3}U_{\beta i}U^{*}_{\beta j}U_{\alpha j}U^{*}_{\alpha i}\exp\left[-i\left(\Phi_{i}-\Phi_{j}\right)\right].
\end{eqnarray}

This tells us that probability depends on the phase differences $\Delta\Phi_{ij}=\Phi_{i}-\Phi_{j}$, it can change the behavior of oscillation.

In flat spacetime, adopting the plane-wave approximation in one spatial dimension, the phase $\Phi_{i}$ for the $i$-th mass eigenstate is expressed as:
\begin{equation}
\Phi_{i}=E_{i}\left(t_{D}-t_{S}\right)-p_{i}\left(x_{D}-x_{S}\right). 
\end{equation}
Clearly, $E_{i}$ and $p_{i}$ are the energy and momentum of the $i$-th mass eigenstate, respectively. For ultra-relativistic neutrinos, where the momentum $p_{i}\approx E_{0}$ (energy of particle at infinity), the phase difference can be approximated using the mass-squared differences $\Delta m^{2}_{ij}=m^{2}_{i}-m^{2}_{j}$.

Finally, we can now write the phase difference in flat spacetime as
\begin{equation}
\Delta\Phi_{ij}=\Phi_{i}-\Phi_{j}\simeq\frac{\Delta m^{2}_{ij}}{2E_{0}}\left|x_{D}-x_{S}\right|. 
\end{equation}
Here $|x_{D}-x_{S}|$ is the beginning distance between the source and detector. This expression can be useful to highlight the dependence of the oscillation probability on the neutrino energy $E_{0}$, the mass-squared differences, and the propagation distance, which are the main parameters in experimental studies on neutrino oscillations.

\section{Phases in Curved Spacetime}

It would be interesting to ask the question: "What is the form of the phase difference in curved spacetime?" To find the phase in curved spacetime, we associate the phase $\Phi_{k}$ with the $k$-th neutrino mass eigenstate that admits a covariant expression, as explained in the pioneering work of Stodolsky~\cite{Stodolsky1979}:
\begin{equation}
\Phi_{k} = \int_{S}^{D} p^{(k)}_{\mu}\, dx^{\mu},
\label{phase_in}
\end{equation}
where the canonical momentum $p^{(k)}_\mu$, conjugate to the coordinate $x^\mu$ for the $k$-th eigenstate, is defined as
\begin{equation}
p^{(k)}_\mu = m_k g_{\mu\nu} \frac{dx^\nu}{ds}.
\label{momentum}
\end{equation}
Here, $m_k$ denotes the invariant mass of the $k$-th eigenstate, and $g_{\mu\nu}$ is the metric tensor of the spacetime. This definition ensures that the dynamics of the particle is consistent with the curved geometry and directly enters the phase integral in \eqref{phase_in}.

The invariant mass squared satisfies the mass-shell condition:
\begin{equation}
m_k^2 = g^{\mu\nu} p^{(k)}_\mu p^{(k)}_\nu,
\label{mass_shell}
\end{equation}
which anchors the propagation of the particle to the spacetime geometry. Substituting \eqref{momentum} into \eqref{mass_shell} shows the consistency of the kinematic description. For geodesics confined to the equatorial plane ($\theta = \pi/2$), the four-velocity component $d\theta/ds = 0$, which leads us to 
\begin{equation}
p^{(k)}_\theta = m_k g_{\theta\nu} \frac{dx^\nu}{ds} = 0.
\end{equation}

Given the invariance of spacetime with respect to $t$ and $\phi$, the momenta $p^{(k)}_t$ and $p^{(k)}_\phi$ are conserved, corresponding to the energy at infinity, $p^{(k)}_t = -E_k$ and the angular momentum, $p^{(k)}_\phi = J_k$. The radial momentum is denoted $p^{(k)}_r = p_k$. In the equatorial plane, the mass-shell condition \eqref{mass_shell} simplifies to
\begin{equation}
m_k^2 = g^{tt} E_k^2 + g^{rr} p_k^2 + g^{\phi\phi} J_k^2,
\label{smash}
\end{equation}
where the absence of cross-terms reflects the diagonal structure of the spacetime. The expression \eqref{smash} is crucial to computing the phase integral \eqref{phase_in}, with momenta $p^{(k)}_t = -E_k$, $p^{(k)}_r = p_k$, $p^{(k)}_\theta = 0$, and $p^{(k)}_\phi = J_k$.

\subsection{Radial Propagation}

We consider the propagation of neutrinos along radial trajectories in a curved spacetime, where the azimuthal angle $\phi$ remains constant ($d\phi = 0$), implying zero angular momentum ($J_k = 0$) for the $k$-th mass eigenstate. This scenario is relevant for neutrinos emitted directly toward or away from a gravitational source along a radial path from the source at radius $r_S$ to the detector at radius $r_D$. The phase $\Phi_k$ is expressed as an integral along the radial null geodesic of a reference particle without mass, and may be written in the following way:
\begin{equation}
\Phi_k = \int_{S}^{D} \left[ -E_k \left( \frac{dt}{dr} \right)_0 + p_k \right] dr.
\label{phase_radial}
\end{equation}
The positions of the source and detector are indicated by $S$ and $D$, respectively.

The differential along the trajectory is derived from the conserved quantities and the metric components $\mathcal{A} = g_{tt}$ and $\mathcal{B} = g_{rr}^{-1}$:
\begin{equation}
\left( \frac{dt}{dr} \right)_0 = \frac{E_0}{p_0} \frac{\mathcal{B}}{\mathcal{A}},
\label{dtdr}
\end{equation}
where $p_0$ is radial momentum of a massless reference particle at infinity.

The radial momentum $p_k$ for the massive neutrino is obtained from the mass-shell condition, is expressed as
\begin{equation}
-m_k^2 = g^{tt} E_k^2 + g^{rr} p_k^2,
\label{massshell}
\end{equation}
as $g^{tt} = \mathcal{A}^{-1}$ and $g^{rr} = \mathcal{B}^{-1}$. By solving \eqref{massshell} for $p_k$, we find
\begin{equation}
p_k = \pm \sqrt{ \frac{\mathcal{B} E_k^2}{\mathcal{A}} - \mathcal{B} m_k^2 },
\label{pk}
\end{equation}

where $m_k$ is the mass of the $k$-th eigenstate. For a massless particle ($m_0 = 0$), the momentum simplifies to
\begin{equation}
p_0 = \pm E_0 \sqrt{ \frac{\mathcal{B}}{\mathcal{A}} }.
\label{p0}
\end{equation}

Substituting \eqref{dtdr}, \eqref{pk}, and \eqref{p0} into the phase integral \eqref{phase_radial}, then we can interpret the phase as
\begin{equation}
\Phi_k = \pm \int_{S}^{D} E_k \sqrt{ \frac{\mathcal{B}}{\mathcal{A}} } \left[ -1 + \sqrt{ 1 - \frac{m_k^2 \mathcal{A}}{E_k^2} } \right] dr.
\label{phase_general}
\end{equation}
To simplify \eqref{phase_general}, we consider the relativistic limit for ultra-relativistic neutrinos ($m_k \ll E_k$), and its energy:
\begin{equation}
E_k \approx E_0 + \mathcal{O} \left( \frac{m_k^2}{2 E_0} \right),
\label{Ek_approx}
\end{equation}
indicating that the energy of the massive neutrino $E_k$ is close to that of the massless reference particle $E_0$ and after considering $\mathcal{AB}=1$ ~\eqref{eq:7} and ~\eqref{eq:8}
\begin{align}
 & \Phi_k \approx \pm\int_{s}^{d} \frac{m_k^2}{2 E_k}\sqrt{\mathcal{AB}} =\pm \frac{m_k^2}{2 E_0} (r_D-r_S). 
 \label{phase_final}
\end{align}

This result arises from expanding the square root term in \eqref{phase_general} and using the approximation \eqref{Ek_approx}. The phase $\Phi_k$ in \eqref{phase_final} is proportional to the radial distance $r_D-r_S$, and the mass squared $m_k^2$, consistent with the phase accumulated in flat spacetime or the same as the weak-field limit of the Schwarzschild spacetime.

\subsection{Non-Radial Propagation in the Equatorial Plane}

Here we investigate the propagation of neutrinos along non-radial trajectories in the equatorial plane of a curved spacetime, where the angular momentum $ J_k$ of the $k$-th mass eigenstate is non-zero $J_k \neq 0$. This scenario is particularly relevant for neutrinos emitted from astrophysical sources, such as supernovae or active galactic nuclei, that traverse regions with significant gravitational curvature. Phase $\Phi_k$ is computed as an integral along a trajectory connecting the source at radius $r_S$ to the detector at radius $ r_D$:
\begin{equation}
\Phi_{k} = \int_{S}^{D} \left[ -E_{k} \left( \frac{dt}{dr} \right)_{0} + p_{k} + J_{k} \left( \frac{d\phi}{dr} \right)_{0} \right] dr.
\label{phase_integral_nonradial}
\end{equation}

The differentials along the trajectory are derived from conserved quantities and metric components $ \mathcal{A} = g_{tt} $, $ \mathcal{B} = g_{rr}^{-1} $, and $\mathcal{D} = g_{\phi\phi} $, which describe the geometry of spacetime. These are expressed as
\begin{equation}
\frac{dt}{dr} = \frac{E_k}{p_k} \frac{\mathcal{B}}{\mathcal{A}}, \quad \frac{d\phi}{dr} = \frac{J_k}{p_k} \frac{\mathcal{B}}{\mathcal{D}}.
\label{diffs_massive_nonradial}
\end{equation}
They become
\begin{equation}
\left( \frac{dt}{dr} \right)_{0} = \frac{E_0}{p_0} \frac{\mathcal{B}}{\mathcal{A}}, \quad \left( \frac{d\phi}{dr} \right)_{0} = \frac{J_0}{p_0} \frac{\mathcal{B}}{\mathcal{D}}.
\label{diffs_massless_nonradial}
\end{equation}

The angular momentum $ J_k $ is related to the geometry of the trajectory through the impact parameter $b$, defined as the perpendicular distance from the center of the gravitational source to the asymptotic trajectory and the velocity at infinity $v_k^\infty$:
\begin{equation}
J_k = E_k b v_k^\infty.
\label{Jk_def}
\end{equation}

In the asymptotically flat region, the velocity of the $k$-th eigenstate is approximated using the relativistic dispersion relation:
\begin{equation}
v_k^\infty = \frac{\sqrt{E_k^2 - m_k^2}}{E_k} \approx 1 - \frac{m_k^2}{2 E_k^2},
\label{vk_approx}
\end{equation}
where $ m_k $ is the mass of the $k$-th eigenstate, and the approximation holds for ultra-relativistic neutrinos ($m_k \ll E_k$). 

Thus, the angular momenta are explicitly given as
\begin{align}
J_k &\approx E_k b \left( 1 - \frac{m_k^2}{2 E_k^2} \right), 
\label{Jk_approx}\\
J_0 &= E_0 b,
\label{J0_def}
\end{align}
where terms up to $ \mathcal{O}(m_k^2 / E_k^2) $ are retained in the relativistic limit. The mass-shell condition for the $k$-th eigenstate in contravariant form is written as
\begin{equation}
-m_k^2 = g^{tt} E_k^2 + g^{rr} p_k^2 + g^{\phi\phi} J_k^2.
\label{mass_shell_nonradial}
\end{equation}

This condition allows us to relate the radial momenta of the massive and massless particles, leading to the relation
\begin{equation}
\frac{p_0 p_k}{E_0 E_k \mathcal{B}} = \frac{1}{\mathcal{A}} - \frac{b^2}{\mathcal{D}} - \frac{m_k^2}{2 E_k^2},
\label{radial_relation_nonradial}
\end{equation}
which holds for both massive ($ m_k \neq 0 $) and massless ($ m_0 = 0 $) particles.
After substituting Eqs.~\eqref{diffs_massive_nonradial}, \eqref{diffs_massless_nonradial}, \eqref{Jk_approx}, and \eqref{radial_relation_nonradial} into Eq.~\eqref{phase_integral_nonradial}, and applying the relativistic approximation $ E_k \approx E_0 $ (which is valid for ultra-relativistic neutrinos), the phase integral simplifies by using the relativistic approximation
\begin{eqnarray}
 &&\Phi_{k} = -\frac{m_{k}^{2}}{2E_{0}}\int_{S}^{D}\frac{E_{0}\mathcal{B}}{p_{0}}dr  \\ &&= \pm\frac{m_{k}^{2}}{2E_{0}}\int_{S}^{D}\sqrt{\mathcal{A}\mathcal{B}}\left(1-\frac{b^{2}\mathcal{A}}{\mathcal{D}}\right)^{-1/2}dr\nonumber \\
&&= \pm\frac{m_{k}^{2}}{2E_{0}}\int_{S}^{D}\left[1-\frac{b^{2}}{r^2}\left(1-\frac{2 M}{r}+\frac{a }{r} \ln\left(\frac{r}{| a| }\right) \right)\right]^{-\frac12}dr. \nonumber 
\label{phase_nonradial_simplified}
\end{eqnarray}

For neutrinos escaping the gravitational potential, we use the weak-field approximation to expand the integrand:
\begin{eqnarray}
\Phi_{k} &=& \pm\frac{m_{k}^{2}}{2E_{0}}\int_{S}^{D}\Bigg[\frac{1}{\sqrt{1-{b^{2}}/{r^{2}}}}-\frac{b^2 M}{r^3 \left(1-{b^2}/{r^2}\right)^{3/2}}\nonumber \\ &&+\frac{a b^2 \ln \left({r}/{| a| }\right)}{2 r^3 \left(1-{b^2}/{r^2}\right)^{3/2}}\Bigg]dr.
\end{eqnarray}

Integration from source $r_S$ to detector $r_D$ yields the following
\begin{eqnarray}
\Phi_{k} &=& \frac{m_{k}^{2}}{2E_{0}} \Bigg[ \frac{M}{\sqrt{1-{b^2}/{r_D^2}}} - \frac{M}{\sqrt{1-{b^2}/{r_S^2}}} \nonumber \\ && + r_D \sqrt{1-{b^2}/{r_D^2}} -r_S \sqrt{1-{b^2}/{r_S^2}} \nonumber \\
&& - \frac{a \ln \left({r_D}/{|a|}\right)}{2 \sqrt{1-{b^2}/{r_D^2}}} + \frac{a \ln \left({r_S}/{|a|}\right)}{2 \sqrt{1-{b^2}/{r_S^2}}} \nonumber \\ && + \frac{1}{2} a \ln \left( \frac{r_D \left( \sqrt{1-{b^2}/{r_D^2}} + 1 \right)}{r_S \left( \sqrt{1-{b^2}/{r_S^2}} + 1 \right)} \right) \Bigg]\nonumber \\
&\approx &\frac{M r_D}{\sqrt{r_D^2 - b^2}} - \frac{Mr_S}{\sqrt{r_S^2 - b^2}} + \sqrt{r_D^2 - b^2} - \sqrt{r_S^2 - b^2}\nonumber \\ && +a \ln\left(\frac{r_D}{r_S}\right) - \frac{a b^2}{4 r_D^2} \ln\left(\frac{r_D}{|a|}\right) + \frac{a b^2}{4r_S^2} \ln\left(\frac{r_S}{|a|}\right) .
\label{eq:42}
\end{eqnarray}

In Eq \eqref{eq:42}, the last two terms become significantly smaller at least $10^{12}$ times compared to the other terms in our case, due to the presence of $r_S^2$ and $r_D^2$ in their denominators.
The expression
\begin{eqnarray}
\Phi_{k} &\approx& \frac{M r_D}{\sqrt{r_D^2 - b^2}} - \frac{Mr_S}{\sqrt{r_S^2 - b^2}} \nonumber \\ && + \sqrt{r_D^2 - b^2} - \sqrt{r_S^2 - b^2}+ a \ln\left(\frac{r_D}{r_S}\right). 
\end{eqnarray}
gives a summary of the phase accumulated by neutrinos traversing non-radial paths in the equatorial plane of a spacetime described by non-zero angular momentum ($ J_k \neq 0 $). The integral accounts for the gravitational effects on the trajectory of the neutrino, with the impact parameter $b$ and the metric components $ \mathcal{A} $, $ \mathcal{B} $, and $ \mathcal{D} $ encoding the geometry of spacetime. The phase $ \Phi_k $ is crucial for understanding neutrino oscillations in curved spacetimes as it influences the interference patterns observed at the detector. 

We now turn to the case of neutrino propagation in the presence of a gravitational lens. In this scenario, the curved geometry bends the neutrino trajectory, producing two distinct segments of the path: one extending from the source $S$ to the point of closest approach $C$, and the other from $C$ to the detector $D$. The accumulated phase $\Phi_k$ along this diverted trajectory is therefore expressed as a sum of two integrals, corresponding to the inbound and outbound paths:
\begin{equation}
\Phi_{k} = \frac{m_{k}^{2}}{2E_{0}}\left[\int_{r_C}^{r_S}\sqrt{\frac{\mathcal{A}\mathcal{B}}{1-\frac{b^{2}\mathcal{A}}{\mathcal{D}}}}dr + \int_{r_C}^{r_D}\sqrt{\frac{\mathcal{A}\mathcal{B}}{1-\frac{b^{2}\mathcal{A}}{\mathcal{D}}}}dr\right].
\label{phase_integral}
\end{equation}

The distance of closest approach $r_C$ is obtained from the condition that the radial derivative of the null trajectory vanishes,
\begin{equation}
\left(\frac{dr}{d\phi}\right)_0 = \frac{p_0(r_C)\mathcal{D}}{J_0\mathcal{B}} = 0.
\label{closest}
\end{equation}
Solving Eq.~\eqref{closest} in the weak-field limit yields an approximate analytical expression for $ r_C $:
\begin{equation}
r_C = \sqrt{b^2+M^2}-M \approx b - M,
\label{rC_approx}
\end{equation}
where $M$ is the gravitational mass of the lensing object, the approximation in Eq.~\eqref{rC_approx} is valid when the impact parameter $b$ is significantly larger than $M$, which corresponds to weak lensing situations.

Performing the integration of Eq.~\eqref{phase_integral}, one obtains the following explicit form for the neutrino phase:
\begin{eqnarray}
\Phi_{k} &\approx& \frac{m_{k}^{2}}{2E_{0}}\Bigg[\sqrt{r_D^2-b^2}+\sqrt{r_S^2-b^2}  \nonumber \\ && + M\left(\frac{b}{\sqrt{r_D^2-b^2}} + \frac{b}{\sqrt{r_S^2-b^2}} \right. \nonumber \\
&& \left. + \sqrt{\frac{r_D-b}{r_D+b}} + \sqrt{\frac{r_S-b}{r_S+b}}\right)  + a\ln\frac{r_Sr_D}{M^2}\Bigg].
\label{phase_full}
\end{eqnarray}

The expression in Eq.~\eqref{phase_full} demonstrates the interplay between the geometric path length, the gravitational potential of the lens and logarithmic contributions that arise from spacetime curvature.

In the regime where the impact parameter is much smaller than the source and detector distances ($ b \ll r_{S}, r_{D} $), Eq.~\eqref{phase_full} simplifies to
\begin{eqnarray}
&&\Phi_{k} \approx \frac{m_{k}^{2}}{2E_{0}}(r_S+r_D)\nonumber \\ && \ \  \times \left[1-\frac{b^2}{2r_Sr_D} + \frac{2M}{r_S+r_D} \right.  \left. + \frac{a}{r_S+r_D}\ln\frac{r_Sr_D}{M^2}\right],
\label{phase_limit}
\end{eqnarray}
which highlights the leading-order effects: a dominant term proportional to the total path length $r_S + r_D $, small geometric corrections from the finite impact parameter $ b $, and gravitational corrections from the lens mass $ M $ and the logarithmic term. 

This result clearly shows how gravitational lensing modifies the accumulated neutrino phase, leading to additional corrections beyond those present in purely radial or non-radial trajectories without lensing. Consequently, gravitational lensing not only deflects the trajectory of neutrinos but also imprints observable modifications on their oscillation phase, which may play a crucial role in astrophysical neutrino detection experiments.

\section{Neutrino Oscillation Probabilities}

The phenomenon of neutrino oscillation is considered a cornerstone of modern particle physics, and it acquires a profound new dimension when considered in curved spacetime, particularly gravitational lensing. Here, we undertake an exploration of the oscillation probabilities for neutrinos travelling multiple paths in the spacetime, where the interplay of quantum mechanics and GR manifests in subtle ways. We aim to compute the probability that a neutrino emitted in a flavor eigenstate at a source $S$ evolves into another flavor eigenstate upon reaching a detector $D$, accounting for the gravitational effects that bend its possible trajectories.

We assume that a neutrino emitted at the source $S$ in a flavor eigenstate $|\nu_{\alpha}, S\rangle$, expressed as a superposition of mass eigenstates: $|\nu_{\alpha}, S\rangle=\sum_{i}U_{\alpha i}|\nu_{i}\rangle$, where $U_{\alpha i}$ are elements of the Pontocorvo-Maki-Nakagawa-Sakata (PMNS) mixing matrix, and $|\nu_{i}\rangle$ denote the mass eigenstates with masses $m_{i}$. As the neutrino propagates through the spacetime, it may follow distinct geodesics, say paths $p$ and $q$ due to gravitational lensing by a massive object, such as a black hole. These paths are characterised by different impact parameters $b_{p}$ and $b_{q}$, introducing distinct phase shifts, leading to interference effects at the detector $D$.
 
The evolved state at the detector is therefore expressed as
\begin{equation}
|\nu_{\alpha}, D\rangle = N \sum_{i} U^{*}_{\alpha i} \sum_{p} \exp\!\left(-i \Phi^{p}_{i}\right) |\nu_{i}\rangle,
\label{detector_State}
\end{equation}
where $\Phi^{p}_{i}$ is the phase accumulated by the eigenstate $i$-th mass $|\nu_{i}\rangle$ along the trajectory $p$, and $N$ is an overall normalization factor that ensures the total probability is conserved.

The transition probability from an initial flavor $\nu_{\alpha}$ to a final flavor $\nu_{\beta}$ is then obtained by taking the squared modulus of the overlap between the evolved state \eqref{detector_State} and the flavor eigenstate $|\nu_{\beta}\rangle$ at the detector:
\begin{eqnarray}
\mathcal{P}_{\alpha\beta} &=& \big|\langle \nu_{\beta} | \nu_{\alpha}, D \rangle \big|^{2}
\nonumber \\ &=& |N|^{2} \sum_{i, j} U_{\beta i} U^{*}_{\beta j} U_{\alpha j} U^{*}_{\alpha i} 
\sum_{p, q} \exp\!\left(-i \Delta\Phi^{pq}_{ij}\right),
\label{osc_prob_general}
\end{eqnarray}
where the phase difference between two contributions is defined as
\begin{equation}
\Delta\Phi^{pq}_{ij} = \Phi^{p}_{i} - \Phi^{q}_{j}.
\label{phase_diff_def}
\end{equation}

The normalization constant is obtained from the condition $\mathcal{P}_{\alpha\alpha} = 1$ when no oscillation occurs. Explicitly, it takes the form
\begin{equation}
|N|^{2} = \left(\sum_{i} |U_{\alpha i}|^{2} \sum_{p, q} \exp\!\left(-i \Delta\Phi^{pq}_{ii}\right)\right)^{-1}.
\label{normalization}
\end{equation}

Equation~\eqref{osc_prob_general} makes explicit the double sum over both the mass indices $(i,j)$ and the different possible lensing paths $(p,q)$. The first summation encodes the standard quantum interference between different neutrino mass eigenstates, while the second arises from the multiplicity of classical trajectories in a curved spacetime due to gravitational lensing. The resulting probability therefore, captures both flavor oscillations and path interference effects, leading to a more complex structure than in flat spacetime, where only the $(i,j)$ sum remains.

\begin{figure*}[t]
\centering
\resizebox{1\textwidth}{!}{
\begin{tikzpicture}
\tikzstyle{every node}=[font=\LARGE]
\draw (6, 7.5) circle (1cm);
\filldraw[black] (6, 7.5) circle (1cm);
\node at (6, 6) {Gravitating object};
\draw[thick] (-2.5, 7.5) .. controls (1, 9) and (3, 10.5) .. (15.5, 11);
\draw [->, >=Stealth] (-2.5, 7.5) -- (16.5, 7.5) node[right] {$x$};
\draw [->, >=Stealth] (6, 7.5) -- (5.75, 15.5) node[above] {$y$};
\draw [short] (6, 7.5) -- (15.5, 11);
\draw [short] (-2.5, 7.5) -- (14.25, 16.75);
\draw [->, >=Stealth, dashed] (6, 7.5) -- (0.25, 16.75) node[left] {$y'$};
\draw [->, >=Stealth, dashed] (6, 7.5) -- (16.5, 13.25) node[right] {$x'$};
\draw [short] (3.85, 11) -- (15.5, 11);
\node at (-2.8, 7.5) {S};
\node[below] at (-2.6, 7.25) {Source};
\node at (16, 11) {D};
\node[above] at (16, 11.25) {Detector};
\node at (3.4, 11.25) {$b$};
\node at (5.6, 11.45) {$\delta$};
\node at (2.5, 6.75) {$r_S$};
\node at (12, 9) {$r_D$};
\node at (5.5, 9.2) {$\varphi$};
\coordinate (A1) at (3.85, 11);
\coordinate (X1) at (15.5, 11);
\coordinate (Y1) at (14.25, 16.75);
\draw[] (X1) -- (A1) -- (Y1)
pic [ draw, angle radius = 1.5 cm] {angle = X1--A1--Y1};
\coordinate (A) at (6,7.5);
\coordinate (X) at (5.75,15.5);
\coordinate (Y) at (5.5,8.3);
\draw[] (X) -- (A) -- (Y)
pic [ draw, angle radius = 1.5 cm] {angle = X--A--Y};
\end{tikzpicture}
}

\caption{\justifying Schematic illustration of weak lensing of neutrinos in the spacetime of a massive object surrounded by DMPF (Adapted from \cite{C5}). Neutrinos propagate from the source {$S$} to the detector {$D$} through the exterior region of the gravitational source. }
\label{fig:1}
\end{figure*}
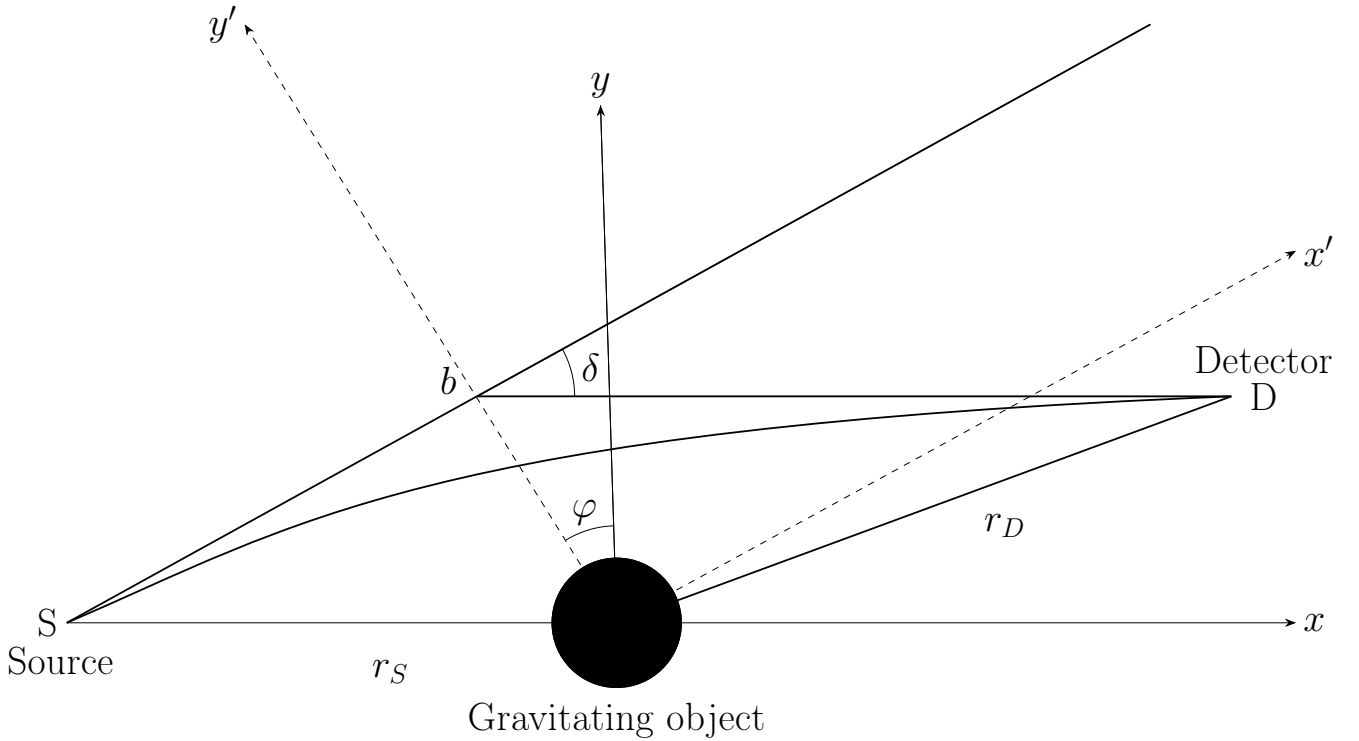

In the presence of gravitational lensing, the phase difference admits a convenient decomposition into two separate contributions, one proportional to the mass-squared difference and the other proportional to the difference in the squared impact parameters:
\begin{equation}
\Delta\Phi^{pq}_{ij} = A_{pq}\, \Delta m^{2}_{ij} + B_{ij}\, \Delta b^{2}_{pq},
\label{phase_split}
\end{equation}
with the notations $\Delta m^{2}_{ij} = m^{2}_{i} - m^{2}_{j}$, $\Delta b^{2}_{pq} = b^{2}_{p} - b^{2}_{q}$, $\Sigma m^{2}_{ij} = m^{2}_{i} + m^{2}_{j}$, and $\Sigma b^{2}_{pq} = b^{2}_{p} + b^{2}_{q}$ these have been introduced for compactness. The explicit forms of the coefficients are given in the form:
\begin{eqnarray}
A_{pq} &=& \frac{r_{S} + r_{D}}{2E_{0}}
\Bigg[1 + \frac{2M}{r_{S}+r_{D}} 
\nonumber \\ && + \frac{a}{r_{S}+r_{D}} \ln\!\left(\frac{r_{S} r_{D}}{M^{2}}\right) 
- \frac{\Sigma b^{2}_{pq}}{4 r_{S} r_{D}}\Bigg], 
\label{A_coeff} \\
B_{ij} &=& -\frac{\Sigma m^{2}_{ij}}{8E_{0}}
\left(\frac{1}{r_{S}} + \frac{1}{r_{D}}\right).
\label{B_coeff}
\end{eqnarray}

\section{Two-Flavor Model}

Now we want to show thatin many astrophysical and gravitational contexts, neutrino oscillations can be straightforward to understand behavior in spacetime within the two-flavor approximation. Although the full description involves three flavors, we carry it out only to show that one can construct a consistent two-flavor framework. Indeed, let us use proper reasoning to note that this holds when $\Delta m_{31}^{2} \gg \Delta m_{21}^{2}$ and $\theta_{13}$ is small.

To compute the oscillation probability for $\nu_e \to \nu_\mu$ in the presence of two gravitationally lensed paths, we calculate the transition amplitude from the source flavor state $\ket{\nu_e, S}$ to the detected flavor state $\ket{\nu_\mu, D}$. The probability is written as 
\begin{align}
\mathcal{P}^{\rm lens}_{e\mu} &= |N|^{2}\sin^{2}2\alpha\Bigg[\sin^{2}\left(\Delta m^{2}\frac{A_{11}}{2}\right)\nonumber \\ & +\sin^{2}\left(\Delta m^{2}\frac{A_{22}}{2}\right)-\cos(\Delta b^{2}B_{12})\cos(\Delta m^{2}A_{12}) \nonumber \\
&+\frac{1}{2}\cos(\Delta b^{2}B_{11})+\frac{1}{2}\cos(\Delta b^{2}B_{22})\Bigg], 
\end{align}
With normalization:
\begin{equation}
|N|^{2}=\frac{1}{2}\left[1+\cos^{2}\alpha\cos(\Delta b^{2}B_{11})+\sin^{2}\alpha\cos(\Delta b^{2}B_{22})\right]^{-1}\ .
\end{equation}

Here, $\Delta m^{2}=m_{2}^{2}-m_{1}^{2}$, $\Delta b^{2}=b_{1}^{2}-b_{2}^{2}$, and $b_{1}$, $b_{2}$ are the impact parameters of the two paths, determined by solving a polynomial equation derived from the lensing geometry in spacetime. Probability is sensitive to neutrino mass ordering and absolute masses via $\Sigma m^{2}_{ij}$. In a numerical example using the Sun-Earth system ($r_{D}=10^{8}\, \mathrm{km}$, $r_{S}=10^{5}r_{D}$, $E_{0}=10\, \mathrm{MeV}$, $|\Delta m^{2}|=10^{-3}\, \mathrm{eV}^{2}$), the interference terms reveal the interplay of quantum mechanics and spacetime geometry, offering a probe of both neutrino properties and the structure of compact objects.

The main point of the present work is a better understanding of the influence of gravitational lensing parameters on neutrino oscillation probabilities; we examine the variations within a realistic astrophysical context. This requires expressing the impact parameter in terms of observable geometrical quantities. For this purpose, we refer to Fig.~\ref{fig:1}, which depicts a schematic representation of weak gravitational lensing in spacetimes. In this adopted configuration, neutrinos are emitted from a source at the point $S$, gravitationally lensed by a massive compact object described by the spacetime surrounded by DMPF, and subsequently detected at the point $D$.

It is shown that in Fig.~\ref{fig:1}, the radial distances from the lensing object to the source and detector can be denoted as $r_S(x, y)$ and $r_D(x, y)$ in the Cartesian coordinate system $\{x, y\}$. Alternatively, we introduce a rotated coordinate system $\{x', y'\}$ obtained by rotating $\{x, y\}$ at an angle $\varphi$, with the transformation equations $x' = x \cos \varphi + y \sin \varphi$ and $y' = -x \sin \varphi + y \cos \varphi$. The deflection angle $\delta$ for this rotated frame is approximated as
\begin{equation}
\delta \approx \frac{y'_D - b}{x'_D} = -\frac{4M}{b} = -\frac{2 R_x}{b}, 
\label{e/deflection}
\end{equation}
where $R_x = 2M$ represents the effective radius and $(x'_D, y'_D)$ denotes the detector's position in the rotated frame. Using the relation $\sin \varphi = b /r_S$, Eq.~(\ref{e/deflection}) can be reformulated to yield
\begin{equation}
(2 R_x x_D + b y_D) \sqrt{1 - \frac{b^2}{r_S^2}} = b^2 \left( \frac{x_D}{r_S} + 1 \right) - \frac{2 R_x b y_D}{r_S}.
\label{e/impact_eq}
\end{equation}

{Solving Eq.~(\ref{e/impact_eq}) provides the impact parameters as functions of $r_S$, $R_x$, and the detector coordinates $(x_D, y_D)$. As an illustrative example, we consider the 
(Sun/
)–Earth system, adopting representative geometrical parameters and modeling the 
gravitational field within the dark matter framework in the background of the phantom field.  Assuming a circular orbit for the detector, we set $x_D = r_D \cos \varphi$ and $y_D = r_D \sin \varphi$. Numerically solving the quartic Eq.(\ref{e/impact_eq}) yields two positive real roots, $b_1$ and $b_2$, for each $\varphi$.}

\begin{figure*}[t]
\includegraphics[width=1\linewidth]{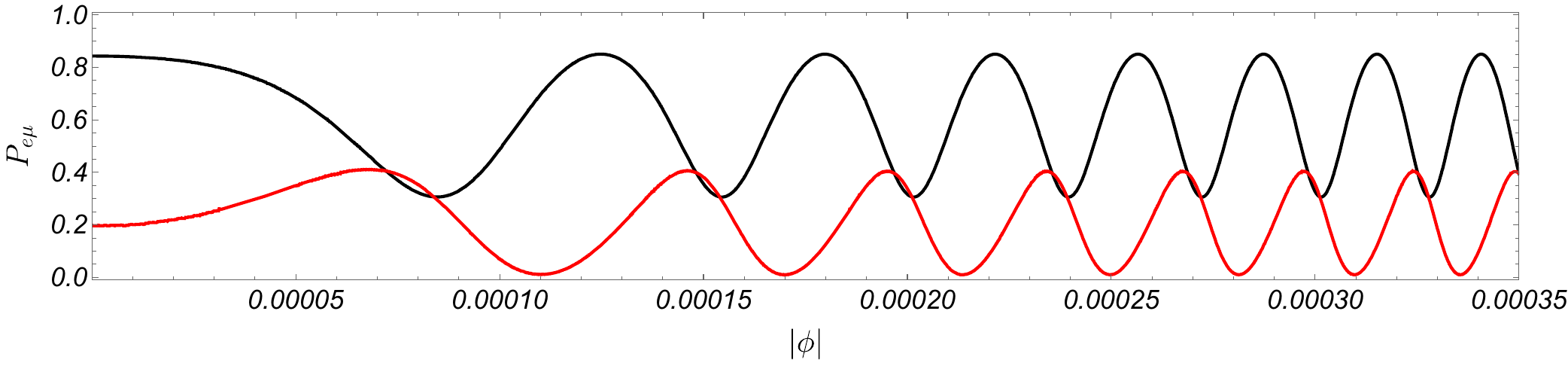}
\includegraphics[width=1\linewidth]{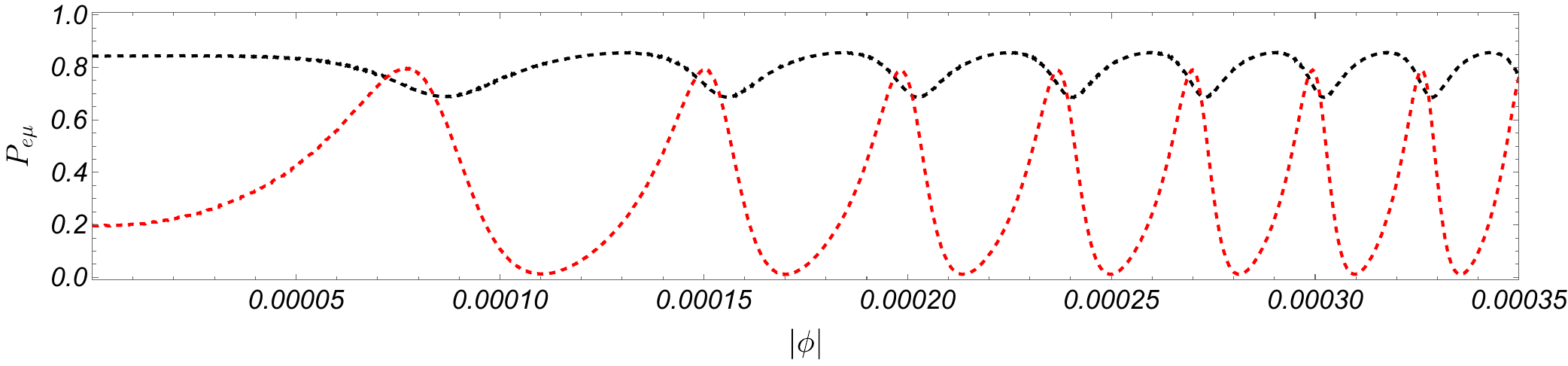} 
\includegraphics[width=1\linewidth]{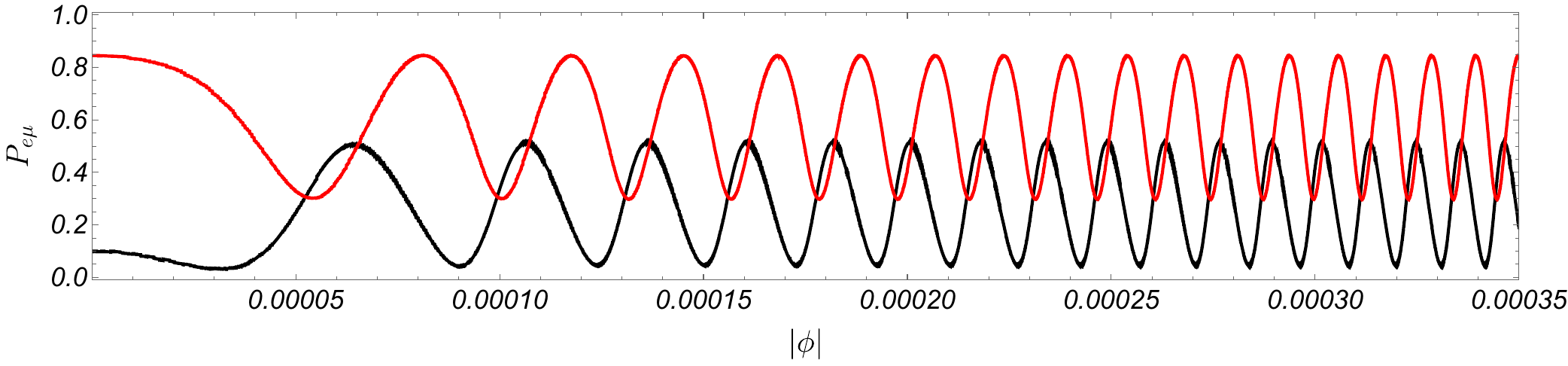}
\includegraphics[width=1\linewidth]{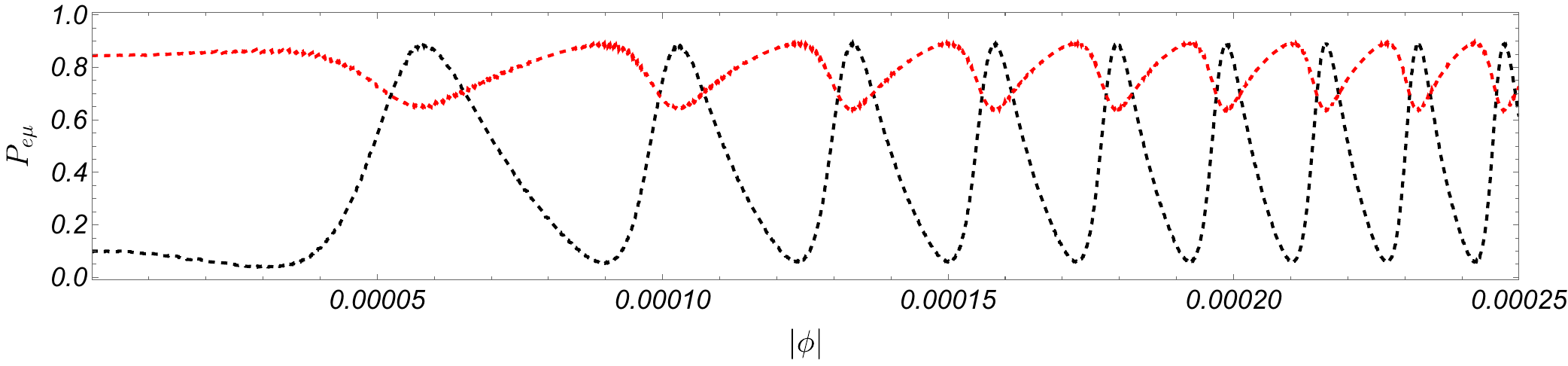} 
\caption{\justifying The neutrino oscillation probability. Black lines indicate the Schwarzschild case ($a/M = 0$), while red lines indicate the dark matter case, with thick lines denoting the normal mass hierarchy (NH, $\Delta m^2 > 0$) and dashed lines denoting the inverted mass hierarchy (IH, $\Delta m^2 < 0$). 
\text{Top two panels:} neutrino oscillation probability including gravitational lensing by Sagittarius A*, with $E_0 = 10^{8}$ MeV and $a/M = 0.15$. 
\text{Bottom two panels:} neutrino oscillation probability for M87*, with a (theoretical) high energy $E_0 = 10^{11}$ MeV and $a/M = 0.335$.
Fixed parameters: mixing angle $\alpha = 33.65^\circ$ \cite{Esteban:2024eli}, mass $M = M_\odot$, squared mass splitting $\Delta m^2 = 10^{-3}\,\text{eV}^2$.}
\label{fig:extra}
\end{figure*}

\begin{figure*}[t]
\includegraphics[width=1\linewidth]{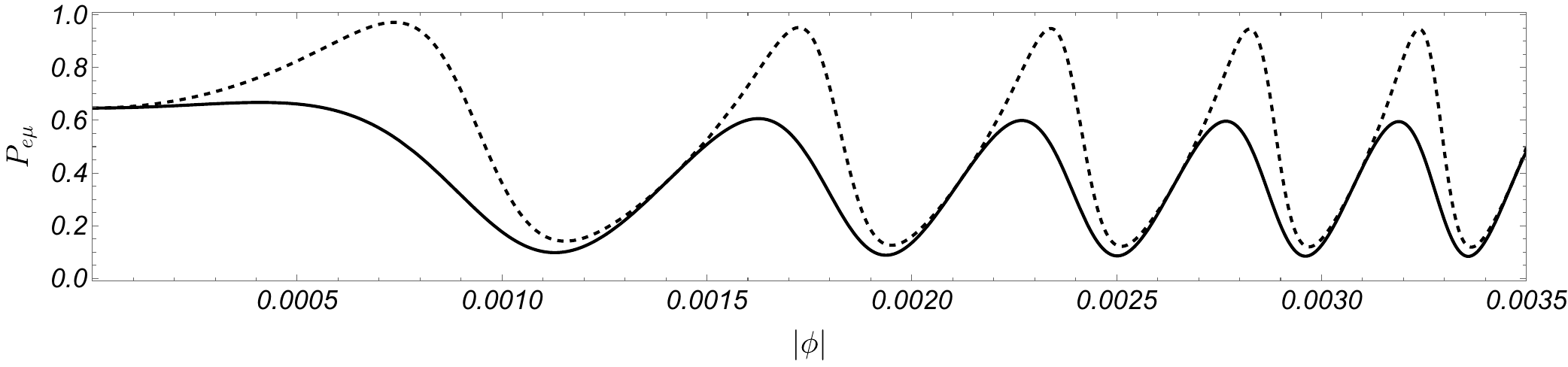}
\caption{\justifying Neutrino oscillation probability for a solar-system-scale black hole, with $E_0 = 10$ MeV and $a/M = 7.5 \times 10^{-12}$. Thick lines denote the normal mass hierarchy (NH, $\Delta m^2 > 0$) and dashed lines denote the inverted mass hierarchy (IH, $\Delta m^2 < 0$). Fixed parameters: mixing angle $\alpha = 33.65^\circ$ \cite{Esteban:2024eli}, mass $M = M_\odot$, squared mass splitting $\Delta m^2 = 10^{-3}\,\text{eV}^2$.}
\label{fig:solar}
\end{figure*}

\begin{figure*}[t]
\includegraphics[width=1\linewidth]{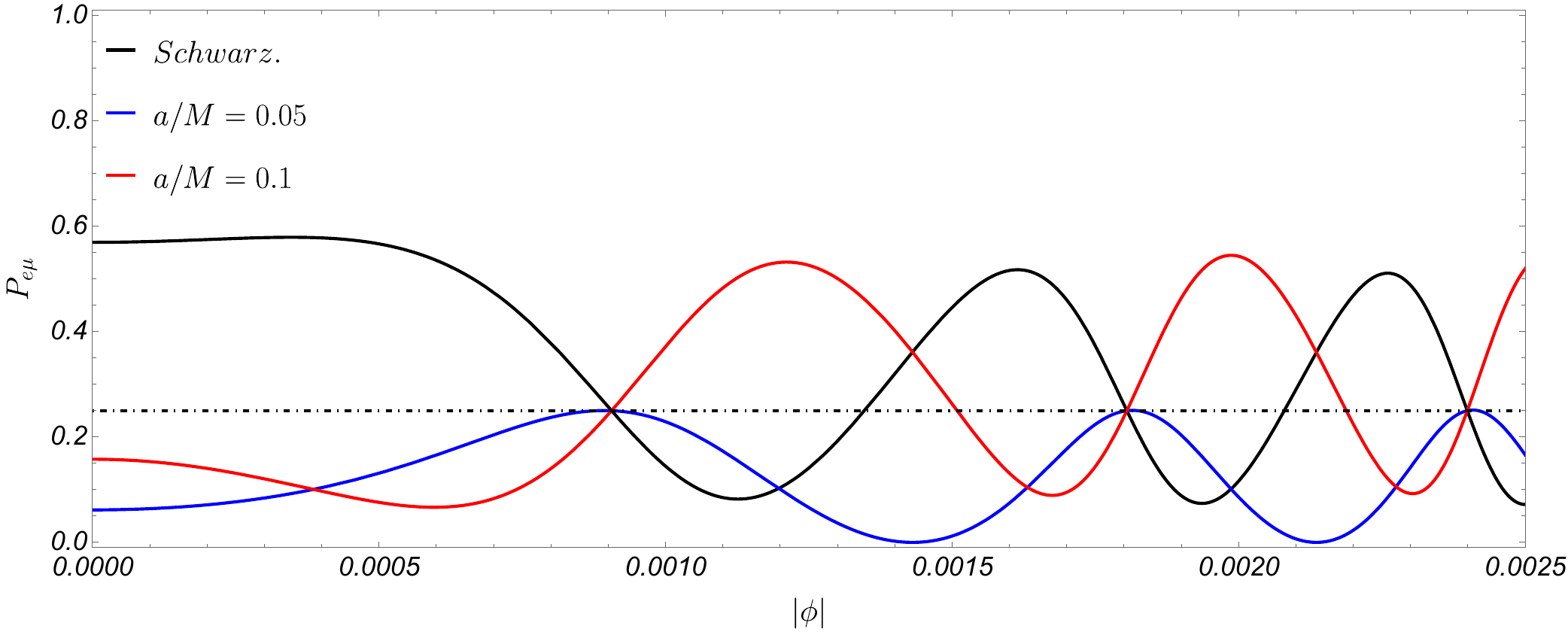}
\includegraphics[width=1\linewidth]{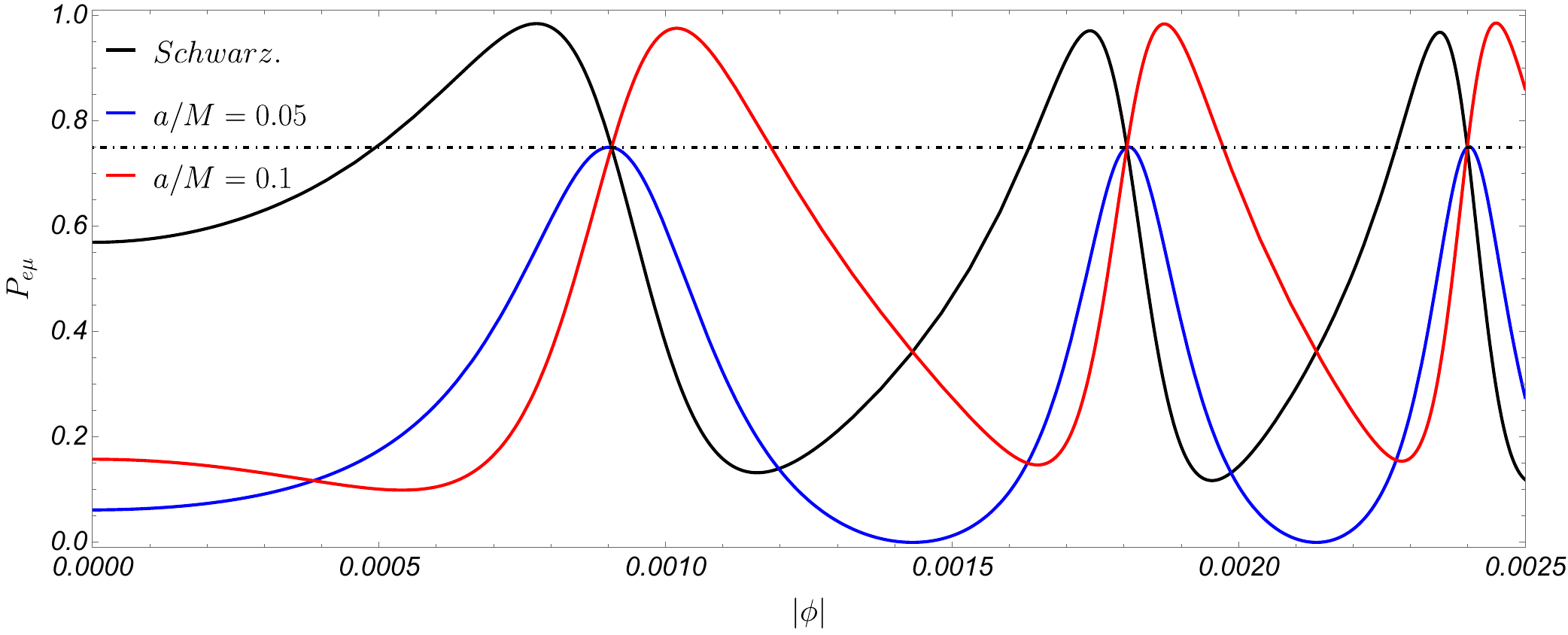} 
 \caption{\justifying The oscillation probability is plotted as a function of the azimuthal angle $\varphi$, highlighting the impact of $a$. 
{Top panel:} Neutrino oscillation probability $P_{\nu_e \to \nu_\mu}(\varphi)$ for Schwarzschild spacetime ($a \to 0$, black solid line) and modified spacetime with dimensionless parameters $a/M = 0.05$ (blue line) and $a/M = 0.1$ (red line), assuming normal mass hierarchy (NH, $\Delta m^2 > 0$). 
{Bottom panel:} Same configuration for inverted mass hierarchy (IH, $\Delta m^2 < 0$). Fixed parameters: mixing angle $\alpha = \pi/6$, mass $M = M_\odot$, squared mass splitting $\Delta m^2 = 10^{-3} {eV^2}$.}
\label{fig:2}
\end{figure*}

Our analysis calculates neutrino oscillation probabilities for impact parameters \(b_p\) that satisfy \(R_x \ll b_p \ll r_D\), thereby validating the weak lensing approximation under conditions where neutrinos travel a considerable distance from the DMPF object. Still, the detector is much farther away than the impact parameter. 

{Additionally, for real-life cases, we have chosen three astronomical objects for checking neutrino oscillations: Solar, Sgr A*, and M87* as lensing objec,ts which have their own dark matter values, see Table~\ref{tab:distances}}

\begin{table}[h]
\centering
\caption{Distances and Masses of Solar, Sagittarius A*, and M87* Systems}
\label{tab:distances}
\begin{tabular}{l l l}
\toprule
\textbf{Object} & \textbf{Quantity} & \textbf{Value} \\
\midrule
\multirow{3}{*}{Solar} 
    & $R_d$ (Sun--Earth) & \SI{1.496e11}{\meter} (1 AU)  \cite{prvsa2016nominal} \\
    & $R_s$ ($\alpha$ Centauri--Sun) & \SI{4.13e16}{\meter}  \cite{Kervella2017} \\
    & Mass & \SI{1.989e30}{\kilogram}  \cite{prvsa2016nominal} \\
\midrule
\multirow{3}{*}{Sgr A*} 
    & $R_d$ (Sgr A*--Earth) & \SI{2.52e20}{\meter} (8178 pc) \cite{Gravity:2019nxk} \\
    & $R_s$  & 5 $R_d$ (Hypotetical source)   \\
    & Mass & $4.297\times10^{6}\,M_\odot$ \cite{Gravity:2019nxk} \\
\midrule
\multirow{3}{*}{M87*} 
    & $R_d$ (M87*--Earth) & \SI{5.18e23}{\meter} (16.8 Mpc) \cite{EventHorizonTelescope:2019pgp} \\
    & $R_s$ & 5 $R_d$ (Hypotetical source)  \\
    & Mass & $6.5\times10^{9}\,M_\odot$ \cite{EventHorizonTelescope:2019pgp} \\
\bottomrule
\end{tabular}
\end{table}

{Unfortunately, no dedicated published constraints on the logarithmic parameter $a$ from Solar System observations have been identified. The most relevant study is Huo \& Liu~\cite{Huo:2026enq}, who constrain a logarithmic correction of this type using Mercury's perihelion precession. Since their numerical bounds are not accessible to us, all we could use of an order-of-magnitude estimate is instead obtained from the independent Solar-System dark-matter density bound
\[
\rho_{\rm DM}\lesssim10^{5}\ {\rm GeV\,cm^{-3}} \quad \text{at } 1\,\mathrm{AU} \quad ~\cite{Frere:2007pi}.
\]
Using $\rho_{\rm total}=a/(8\pi r^{3})$ with restoring $G,c$,
\[
a=8\pi\frac{G}{c^{2}}\,r^{3}\rho_{\rm phys}.
\]
With $r=1\,\mathrm{AU}=1.496\times10^{11}\,\mathrm{m}$ and $\rho_{\rm phys}=1.78\times10^{-16}\,\mathrm{kg\,m^{-3}}$,
\[
0\leq\frac{a}{M}\lesssim7.5\times10^{-12}.
\]
Meanwhile this parameter for Sgr A*($M=4.0\times10^{6}M_{\odot}$) obtained
by treating it as a cloud-of-strings-type constraint from EHT shadow observations \cite{Anjum:2023axh}, the reported bounds are
\[
0\leq\frac{a}{M}\leq
\begin{cases}
0.0507\text{--}0.0611, & 1\sigma,\\[2mm]
0.1282\text{--}0.1489, & 2\sigma,
\end{cases}
\]
where the range reflects the unknown observer inclination $\theta_o\in[0^\circ,90^\circ]$; the lower $2\sigma$ limit, $a/M=0.1282$, corresponds to the face-on configuration ($\theta_o=0^\circ$).
For M87* ($M=6.5\times10^{9}M_{\odot}$), The corresponding bounds are
\[
0\leq\frac{a}{M}\leq
\begin{cases}
0.0792, & 1\sigma,\\[2mm]
0.3349, & 2\sigma.
\end{cases}
\]}

{After choosing those enormous parameters ($M_{\odot}, R_s, R_d$), we have to choose a proper energy range for neutrinos: Solar neutrinos span roughly 0.1--18 MeV, produced by the pp-chain (about 99\% of the Stars's energy output) and the subdominant CNO cycle in its core \cite{agostini2019,BOREXINO:2020aww}. Both components have been successfully measured by the Borexino experiment. Turing toward the Sgr A* at far higher energy levels ($10^{8}$--$10^{10}$ MeV), IceCube's diffuse astrophysical neutrino flux is thought to originate from powerful cosmic accelerators such as active galactic nuclei, gamma-ray bursts, supernova remnants, pulsars, and black holes \cite{IceCube:2013low}. Confirmed point-source contributors include the blazar TXS 0506+056 \cite{IceCube:2018dnn} and the Seyfert galaxy NGC 1068 \cite{IceCube:2022der}. At the end of the spectrum ($\sim$0.1-3 EeV), cosmogenic (``GZK'') neutrinos are predicted to form when ultra-high-energy cosmic rays coll
ide with cosmic microwave background photons over cosmological distances \cite{Ahlers:2012rz}. Unlike the solar and astrophysical populations, this cosmogenic population remains completely theoretical and has not yet been confirmed as a distinct observational signature, but it could be a perfect candidate for the M87* case.}

{In Fig.~\ref{fig:extra}, we calculated the probability of neutrino oscillation as a function of the lensing at azimuthal angle $\varphi$ by Sgr A* and M87*. They show us how dark matter parameter and the mass-squared difference $|\Delta m^2|$ influence the probability. However, because of the very small value of $a$ in the solar system, it is impossible to see dark matter effects in the neutrino oscillation, see Fig.~\ref{fig:solar}. Therefore, due to a lack of evidence, we continue our calculation with the toy model for pedagogical purposes with imaginary scenery (Solar).}

{The lens mass $M = 1 M_\odot$ and the mass-squared difference $|\Delta m^2| = 10^{-3}$ eV$^2$ are two more model parameters. These values are chosen to show how things work. In the Sun–Earth system, the detector is positioned at Earth's orbital distance, $r_D = 10^5$ km, while the source is located far beyond the Sun at $r_S = 10^5 r_D$. The source emits relativistic neutrinos with an energy $E_0 = 10$ MeV.}

We present the oscillation probabilities of the two-flavor neutrino toy model in Fig.~\ref{fig:2}. Our primary objective is to investigate the sensitivity of these probabilities to the dark sector density parameter as spacetime parameters. In Fig.~\ref{fig:2}, we plot the transition probability $P_{\nu_e \to \nu_\mu}$ as a function of the azimuthal angle $\varphi$. The results are shown for three configurations: $a = 0$ (solid black line, corresponding to the Schwarzschild limit), $a/M = 0.05$ (blue line) and $a/M = 0.1$ (red line). The top panel corresponds to the normal mass hierarchy (NH, $\Delta m^2 > 0$), while the bottom panel displays the inverted mass hierarchy (IH, $\Delta m^2 < 0$). From these graphs, we observe that the oscillation probability is highly sensitive to the deformation parameter of the dark sector density, with the mixing angle fixed at $\alpha = \pi/6$. 

\begin{figure*}
\begin{minipage}[t]{0.481\linewidth}
\includegraphics[width=\linewidth]{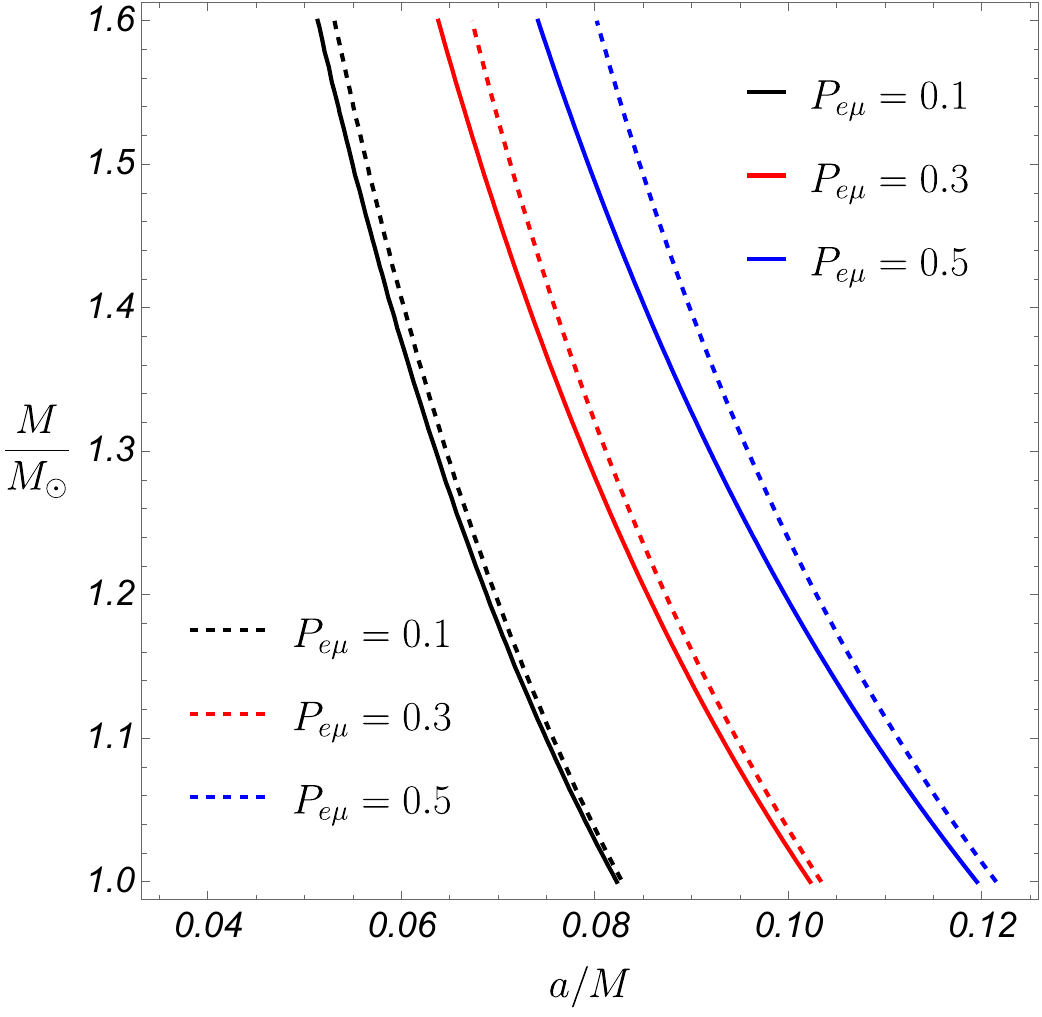}
\caption{\justifying The degeneracy between the mass parameter $M$ and the parameter $a/M$ for a given probability $P_{e\mu}$ is illustrated by the 2D contour plot of the implicit function $M(a/M)$ obtained from $P_{e\mu}(M, a/M) = \mathrm{const}$, with $P_{e\mu} = 0.2$ (black), $0.3$ (red), and $0.5$ (blue). Same configuration for the inverted mass hierarchy (IH, $\Delta m^2<0$) depicted using dashed curves. All panels assume a fixed mixing angle $\alpha = \pi/6$ and squared mass splitting $\Delta m^2 = 1\times 10^{-3}~\mathrm{eV}^2$.}
\label{fig:3}
\end{minipage}\hfill%
\begin{minipage}[t]{0.497\linewidth}
\includegraphics[width=\linewidth]{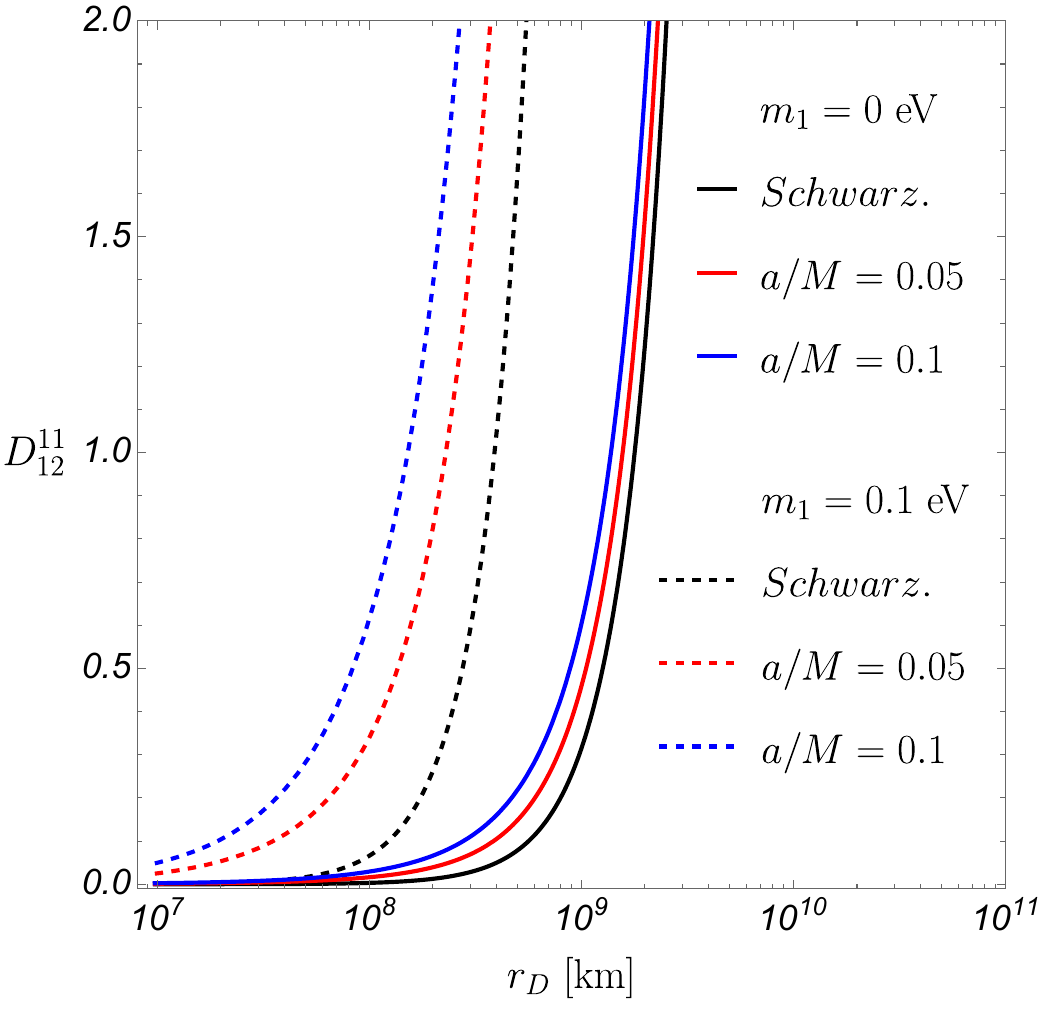}
\caption{\justifying The damping factor $ D^{11}_{12} $ as a function of $ r_D $ for different values of the parameter $ a/M $ with Schwarzschild case ($ a/M = 0 $, black), $ a/M = 0.05 $ (red), and $ a/M = 0.1 $ (blue). Solid lines correspond to $ m_1 = 0~\mathrm{eV} $, while dashed lines correspond to $ m_1 = 0.1~\mathrm{eV} $.}
\label{fig:4}
\end{minipage}%
\end{figure*}

We have also investigated the occurrence of parameter degeneracy, where different combinations of dark sector parameters can lead to the same oscillation probability. This degeneracy is illustrated in Fig.~\ref{fig:3}. We plot the implicit relation $M(a/M)$ obtained for fixed transition probability $P_{e\mu}$: $P_{e\mu} = 0.2$ (black), $0.3$ (red), and $0.5$ (blue). For dashed lines, the same configuration for the inverted hierarchy is represented by dashed curves. In order to highlight their differences, it has been shown that for a given range of dark sector density parameters to get the same probability for IH, we would need slightly larger parameters of $a/M$ than the parameters of NH. Its slight difference becomes larger when we consider greater masses. For all of them, we fix the mixing angle at $\alpha = \pi/6$ and the squared mass splitting at $\Delta m^2 = 10^{-3}~\mathrm{eV}^2$. 

\section{Neutrino Decoherence}

In the previous consideration of neutrino oscillations, the plane waves assumption is used for calculations, which is described as Gaussian wave packets. In reality, we do not fully know how neutrinos propagate, but it is more accurate to imagine them as wave packets rather than as perfect plane waves. Then, taking into account the spatial separation of the different mass states, the decoherence of neutrinos should be considered. In order to quantify this effect, one introduces the characteristic distance called the decoherence length is employed to describe the damping of the oscillation phase between the mass states, which plays a nontrivial role in the propagation of neutrinos at the large astrophysical scale. 

In a gravitational field, such as that near a compact object like a black hole or neutron star, the proper time (the time experienced by a traveling particle) between two points with a fixed proper distance (the actual spatial separation) is shorter than in flat spacetime with no gravity. This happens because gravity warps spacetime, slowing down time slightly. As a result, neutrino wave packets, which rely on proper time to maintain their quantum coherence for oscillations, must travel a greater physical distance in curved spacetime to accumulate the same amount of proper time before they lose coherence, compared to a flat spacetime.

For experimental detection, the coherence is very important because oscillations can only be seen if the neutrino wave packets remain coherent when they reach the detector. As we said before, in the curved spacetime, the altered proper time changes how phases accumulate, so detectors must be within a distance where the wave packets still overlap. If coherence is lost before detection, the oscillation pattern disappears, and we can only observe an averaged flavor mix.

We analyze the impact of gravitational lensing by a black hole on the decoherence length of neutrinos. We imagine that neutrinos are modeled as Gaussian wave packets rather than plane waves, which leads to a modification of the oscillation probability \cite{Swami2021}:
\begin{equation}
P_{\alpha\beta}=\frac{\sum\limits_{ij}U^{*}_{\beta i}U_{\alpha i}U_{\beta j}U^{*}_{\alpha j}
\sum\limits_{m, n}e^{-i\Phi^{mn}_{ij}}e^{-D^{mn}_{ij}}}
{\sum\limits_{i}U_{\alpha i}U^{*}_{\alpha j}\sum\limits_{m, n}e^{-i\Phi^{mn}_{ii}}e^{-D^{mn}_{ii}}}, 
\label{pro}
\end{equation}
where the phase factor $\Phi^{mn}_{ij}$ is given by
\begin{equation}
\Phi^{mn}_{ij}=\left(\Phi^{m}_{i}-\Phi^{n}_{j}\right)-
\frac{\overline{\sigma}^{2}}{\sigma_{D}^{2}}
\left(\vec{p}^{\, D}-\vec{p}^{\, S}\right)\cdot
\left(\vec{X}^{m}_{i}-\vec{X}^{n}_{j}\right),
\end{equation}
and the original damping factor $X^{mn}_{ij}$ is defined as
\begin{equation}
X^{mn}_{ij}=\frac{1}{2}\overline{\sigma}^{2}
\left(\left|\vec{X}^{m}_{i}\right|^{2}+\left|\vec{X}^{n}_{j}\right|^{2}\right), 
\quad \vec{X}^{m}_{i}=\partial_{p}\Phi^{m}_{i}.
\end{equation}

The effective damping factor $D_{ij}$ appearing in Eq.~\eqref{pro} is determined 
by the difference between $X^{mn}_{ij}$ and its minimum value $X^{\hat{m}\hat{n}}_{\hat{i}\hat{i}}$:
\begin{equation}
D^{mn}_{ij}=X^{mn}_{ij}-X^{\hat{m}\hat{n}}_{\hat{i}\hat{i}}, 
\label{e/damp}
\end{equation}

\begin{figure*}[t]
\includegraphics[width=0.8\linewidth]{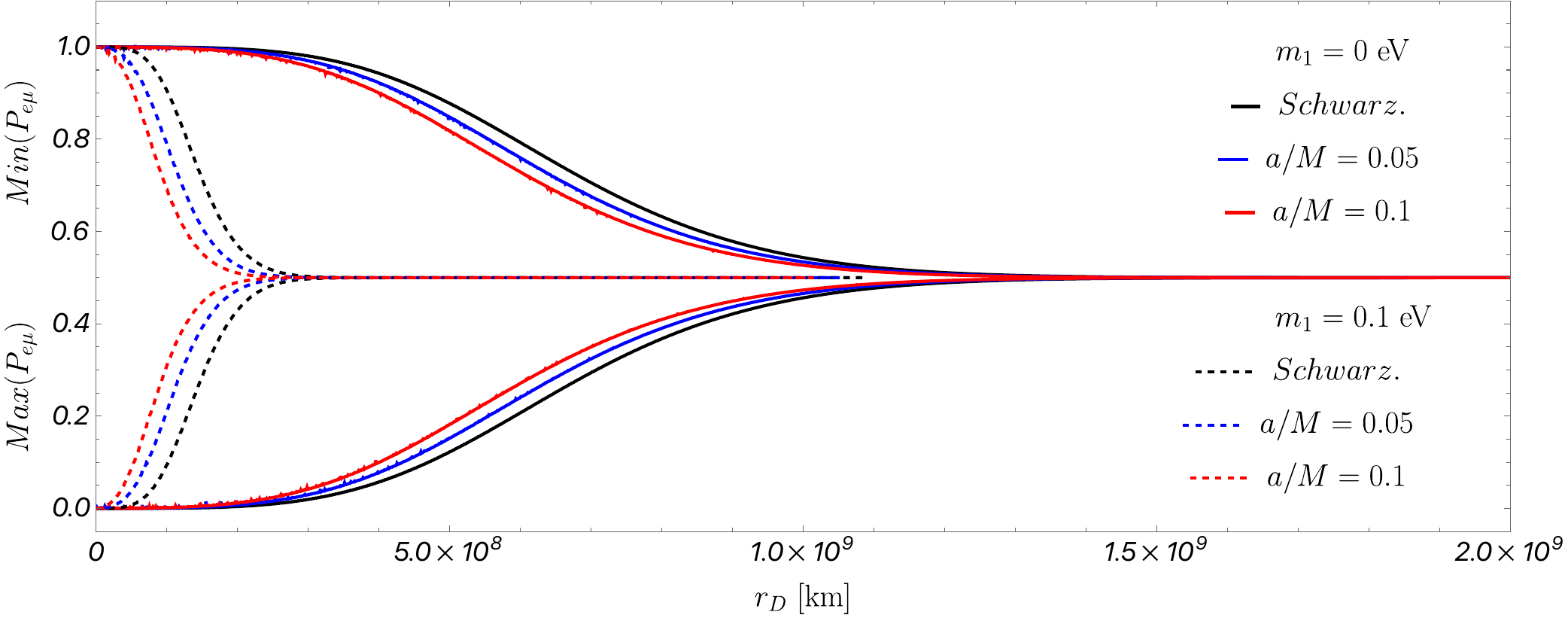}
\caption{\justifying Maximum and minimum transition probability as a function of the detector distance $r_{D}$ for different values of the dark sector density parameter $a/M$. The three panels correspond to Schwarzschild (black), $a/M = 0.05$ (blue), and $a/M = 0.1$ (red). Solid curves represent the case $m_{1} = 0~\mathrm{eV}$, while dashed curves correspond to $m_{1} = 0.1~\mathrm{eV}$. The plots illustrate that while the absolute neutrino mass strongly affects the coherence length, the influence of the dark sector density parameter $a/M$ remains subdominant.}
\label{fig:5}
\end{figure*} 
Here, indices with a hat ($\hat{i}$) correspond to the configuration that minimizes the damping factor. For the two-flavor case, with neutrino trajectories confined to the equatorial plane $\theta=\pi/2$, the path indices are $m, n = 1, 2$ and the mass eigenstate indices are $i, j = 1, 2$. By imposing the ordering $b_{1}\leq b_{2}$ and $m_{1}<m_{2}$, we identify 
$X^{\hat{m}\hat{n}}_{\hat{i}\hat{i}}=X^{11}_{11}$. 
The combined width of the momentum distribution is given by
\begin{equation}
\overline{\sigma}^{2}=\frac{\sigma_{D}^{2}\sigma_{S}^{2}}
{\sigma_{D}^{2}+\sigma_{S}^{2}},
\end{equation}
where $\sigma_{S}$ and $\sigma_{D}$ denote the standard deviations of the momentum distribution functions at the source and the detector, respectively.

For neutrinos propagating non-radially in the presence of a black hole lens, Eq.~\eqref{e/damp} yields the following expression:
\begin{align}
|\vec{X}_i^p|^2 &\simeq 
-\frac{m_i^4}{4E_{\text{loc}}^4\mathcal{A}(r_S)}(r_S + r_D)^2 \notag\\
&\quad\times
\left(1 - \frac{b_p^2}{r_S r_D}
+ \frac{2M}{r_S + r_D}
+ \frac{a}{r_S + r_D}\ln\frac{r_S r_D}{M^2}\right)^2 \notag,\\
\end{align}
where $E_{loc}$ is the neutrino energy observed locally at the source \cite{Swami2021},
\begin{equation}
E_{loc}(r_{S}) = \frac{E_{0}}{\sqrt{-\mathcal{A}(r_{S})}}.
\end{equation}
Consequently, the effective damping factor takes the form.
\begin{align}
D_{ij}^{pq} \approx &-\frac{\bar{\sigma}^{2}(r_{S}+r_{D})^{2}}
{8E_{loc}^{4}\mathcal{A}(r_{S})}\left(1 + \frac{4M}{r_{S}+r_{D}} 
+ \frac{2a}{r_{S}+r_{D}}\ln\frac{r_{S}r_{D}}{M^{2}}\right) \nonumber \\
&\times\bigg[m_{1}^{4}\left(1-\frac{b_{2}^{2}}{r_{S}r_{D}}\right) 
\nonumber  + m_{2}^{4}\left(1-\frac{b_{2}^{2}}{r_{S}r_{D}}\right)-
\\& - 2m_{1}^{4}\left(1-\frac{b_{1}^{2}}{r_{S}r_{D}}\right)\bigg].
\end{align}

To provide a realistic estimate of the decoherence length, we consider a Sun-Earth-based lensing configuration, its relevant parameters being chosen as the source size $R_{x}=3$ km, the local neutrino energy $E_{\mathrm{loc}}=10$ MeV, and the source distance $r_{S}=10^{5}r_{D}$. For simplicity, we restrict our attention to the case in which the source, lens, and detector are aligned~\cite{Swami2021}. The analysis is performed for two different values of the lightest neutrino mass $m_{1}$, while keeping the mass-squared difference fixed at $\Delta m^{2}_{21}=m^{2}_{2}-m^{2}_{1}=10^{-3}\,\mathrm{eV}^{2}$.

The results are summarized in Fig.~\ref{fig:4}, where the damping factor $D^{12}_{11}$ is plotted as a function of the detector distance $r_{D}$. The dashed lines illustrate the effect of the dark sector density parameter by showing the damping factor for $m_{1}=0$ eV with $a/M=0$ (black), $a/M=0.05$ (red), and $a/M=0.1$ (blue). The dashed ones present the same comparison for $m_{1}=0.1$ eV. These results clearly demonstrate that while the absolute neutrino mass has a strong influence on decoherence, the deformation parameter $a/M$ introduces only subdominant corrections. However, in Fig.~\ref{fig:4} the influence of $a/M$ becomes more appreciable on larger mass scales. Thus, the neutrino mass scale still remains the leading factor in determining the decoherence length. The maximum and minimum transition probability envelopes can be examined as a function of the detector distance $r_{D}$ for different values of dark sector density parameter $a/M$, as shown in Fig.~\ref{fig:5}. We select the highest and lowest probability values within an interval $\Delta r_{D}$, and the distance is varied in the range $r_{D} \in [10^{8}, 2\times10^{9}]~\mathrm{km}$, with $\Delta r_{D} = 2 \times 10^{6}~\mathrm{km}$, and the mixing angle is fixed at $\alpha = \pi/4$. The figure shows that neutrino oscillations lose their coherence after a certain propagation scale. The spacetime parameter $a/M$ has a negligible effect on the transition probability once this decoherence regime is reached. However, the coherence length is significantly more affected by the absolute neutrino mass.

\section{Conclusion}

In this work, we examined how the DMPF around a black hole affects neutrino flavor oscillations. Using a two-flavor approximation, we derived the neutrino propagation phase for both purely radial paths and gravitationally lensed paths. For radially moving neutrinos, the main gravitational effects cancel out. As a result, the oscillation phase increases with distance and mass-splitting, just like in flat space. However, neutrinos on deflected trajectories get extra phase shifts. From the analytical calculations, we have seen that the dark matter parameter enters a small correction in the coefficient of the mass-difference term and as an additive contribution to the total phase.

Our numerical analysis of the oscillation probability has confirmed the validity of these trends. Additionally, we examined how the transition probability curves vary with changes in the source angle when the dark matter parameter is modified. We also discovered a degeneracy between the lens mass and the dark matter parameter, indicating that various combinations of these factors can yield the same flavor probability. This suggests that to obtain a specific oscillation probability, one might require a slightly higher dark sector density parameter if the neutrino mass hierarchy is inverted compared to when it is normal.

We further explored the effect of decoherence by modelling neutrinos as Gaussian wave packets. In a gravitational field, proper time dilates, meaning that neutrinos must travel a longer distance to lose coherence. In summary, once decoherence occurs, the dark matter background has minimal impact, and the absolute neutrino mass is the key quantity determining how far oscillations last.

These results can summarize that neutrino oscillations in curved spacetime are sensitive to the gravitational field of dark matter. This in turn suggests that future neutrino astronomy might provide additional insight into dark matter effects in extreme environments.

\begin{acknowledgments}
This research was funded by the National Natural Science Foundation of China (NSFC) under Grant No. U2541210. 
\end{acknowledgments}

\bibliography{bib}
\bibliographystyle{apsrev4-1}
\end{document}